\documentclass{svproc}
\usepackage[utf8]{inputenc}

\usepackage{url}

\usepackage{natbib}
\bibpunct{(}{)}{;}{a}{}{,}%
\usepackage{graphicx}
\usepackage{amsmath,amssymb}
\usepackage{multirow}
\usepackage{longtable}

\usepackage[b5paper]{geometry}
\begin{document}
	\mainmatter              
	\title{Sites of star formation in the tidal structures}
	\titlerunning{Star formation in interacting galaxies}  
	%
	\author{Anatoly Zasov\inst{1,}\inst{2} \and Anna Saburova\inst{1,3} \and
		Oleg Egorov\inst{1} }
	\authorrunning{Anatoly Zasov et al.} 
	%
	\tocauthor{Anatoly Zasov, Anna Saburova, Oleg Egorov}
	\institute{Sternberg Astronomical Institute, Moscow M.V. Lomonosov State University, Universitetskij pr., 13,  Moscow, 119234, Russia,\\
		\email{zasov@sai.msu.ru},\\ 
		\and
		Faculty of Physics, Moscow M.V. Lomonosov State University, Leninskie gory 1,  Moscow, 119991, Russia
		\and
		Institute of Astronomy, Russian Academy of Sciences, Pyatnitskaya st., 48, 119017 Moscow, Russia}
	
	\maketitle

\begin{abstract}
	We give a short review of the current results of studying  of  star formation sites observed  beyond the main discs of galaxies in different interacting systems: Arp~270, Arp~194, Arp~305, NGC~4656 and NGC~90. The  observations were carried out at the 6-meter telescope BTA in SAO RAS with the SCOPRPIO-2 spectrograph. The properties of star forming islands, their mass, dynamics, chemical abundances, their possible fate and the mechanisms inspiring star formation, appear to be different in different systems.   
\end{abstract}

\paragraph{{\bf Introduction}}

A tight interaction  of  gas-rich galaxies is often accompanied by the formation of star-forming   regions outside  galactic discs or at their distant peripheries, in the local areas of the enhanced density of gas.  Dynamical properties, a volume density or a chemical abundance  of gas in these star forming islands may differ significantly from those observed in  spiral arms of galaxies. The mechanism of their formation is an open question. A study of  foci of star formation in tidal structures  is important for understanding the conditions leading to the birth of stars and the evolution of young stellar complexes beyond stellar discs.
 
 A gas associated with star formation in tidal structures, was torn away from parent galaxies, in some cases along with the disc stars.  The average density of the ejected gas should inevitably be very low, and the presence of  the  local dense regions giving birth to stars can be attributed  either to the self-gravity of gas with a low  velocity dispersion in a tidal tail,  or with the processes of interaction of  the expelled gas with the gas environment (a static or dynamic pressure).  
 
A fate of  star-forming islands outside the discs is also not clear. They have a high chance of falling back into a parent galaxy, but in some cases they may remain as long-lived  tidal dwarfs. The distinguishing properties of tidal dwarfs should be a low, if any, content of dark matter, and the moderate gas metallicity, because they consist of gas ejected from the peripheral regions of discs of  more massive parent galaxies.

 Below we give a  short presentation of the main results of study  of  several interacting systems obtained in the frame of  observational program which is carrying out by participants from SAI MSU and SAO RAS.  All spectral observations were performed in the prime focus of the Russian 6-m telescope with SCORPIO-2 spectrograph \citep{Afanasiev2011} in the long slit mode.  In every case a slit was chosen to cross the star-forming regions. We analyzed a distribution of line of-sight velocity for different spectral lines along a slit, together with flux ratios of emission lines and the gas metallicity estimated by different methods. A special attention was paid to the excitation mechanism of emission (a classical HII region or a diffuse ionized gas (DIG)). Using a photometric data, we also tried to estimate the age and the history of star formation in the local star-forming islands.

 A more detailed description of observations, a procedure of data processing, and the  results of investigations of concrete galaxies one may find in the following papers: \citet{zasov2015,zasov2016,zasov2017,zasov2018,zasov2019,zasov2020}.

\paragraph{{\bf 1. System Arp~270}}

Arp~270 = NGC~3395/96  is a tightly
interacting pair of galaxies of comparable luminosity:
a spiral galaxy NGC~3395 and irregular galaxy (Irr
or Sm) NGC~3396, strongly inclined to the line of sight. The
relative motion of galaxies seems to occur in a plane perpendicular
to the line of sight, because their  central systemic 
velocities nearly coincide.
There are numerous separate islands of star formation beyond the main body of galaxies. The most notable of them is the oblong emission region of about one kpc-size between galaxies. 
\begin{figure}
\vspace{-0.5cm}
\centering
	\includegraphics[width=0.40\columnwidth]{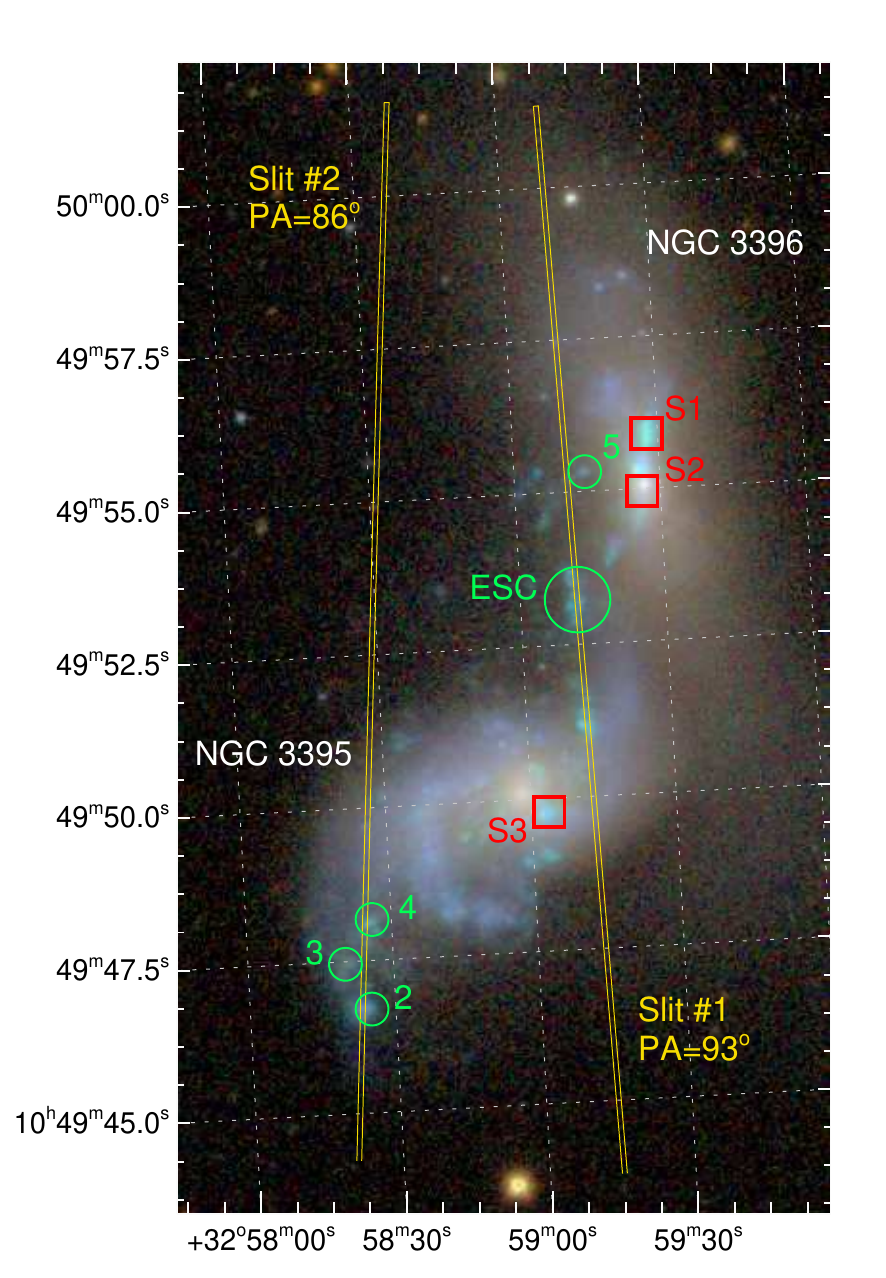}
	\includegraphics[width=0.43\columnwidth]{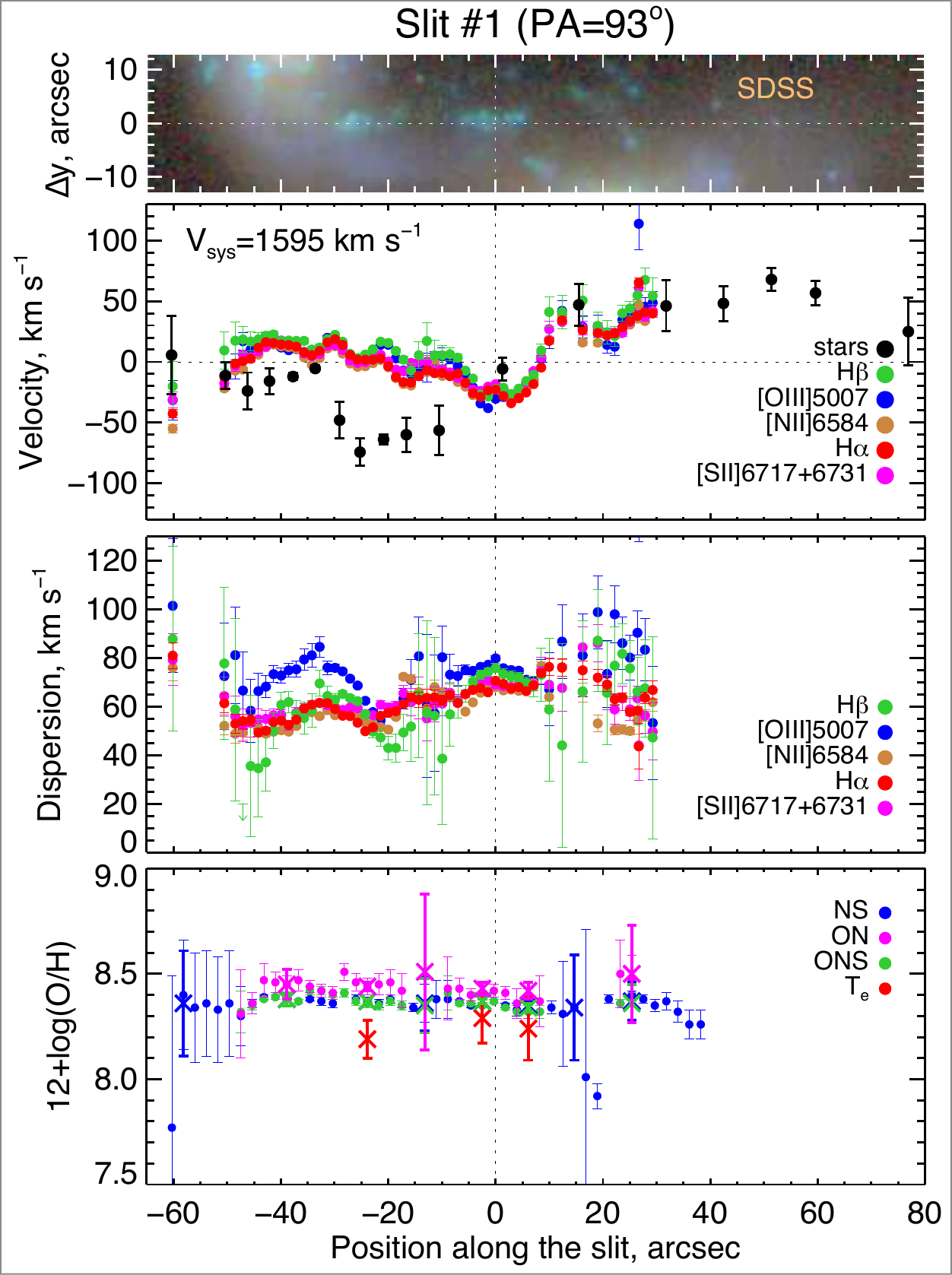}
		\caption{Right:  positions of the slits  overplotted on the {\it g,r,i} ~SDSS image of Arp~270. Left: the  profiles along a slit of stellar and ionized gas kinematics and oxygen abundance of Arp~270 for PA=93.  }
	\end{figure}

We measured the line-of-sight velocities of gas and stars
and chemical abundances distributions along the two slits
crossing the peripheral regions of galaxies (Fig.1).
Irregular velocity distribution and a high velocity dispersion
of emitting gas  exceeding 50 km/s  give evidence of the local non-circular gas flows within a scale of about 1 kpc. 
There is a steep velocity gradient in the region between
galaxies  accompanied by gradual grow of velocity dispersion of gas to the periphery of galactic discs
–   most probably as the result of direct collision of peripheral gaseous 
systems of two galaxies. Note that the brightest emission
region looking as the isolated site of star formation, lies at the beginning of the transition region between galaxies.

The main feature of the  oxygen abundance distribution
is the absence of significant gradients of the ratios (O/H)   along the slits estimated
  by different methods. It evidences the efficient   gas mixing which may be a result of previous   convergence(s) of galaxies.

A comparison of colour indices (ugr) of several discrete sites of star formation beyond the inner regions of the galaxies estimated 
from the SDSS data using  Starburst99 models of stellar evolution  confirmed the young age (T$\le 10^7$
yr) of stellar population without the noticeable substrate of the old population. The extended kpc-size island of star formation  mentioned above does not stand out by its kinematics or abundances, so it hardly
may be considered as the tidal dwarf candidate (hereafter TDG). Its location
at the beginning of transition zone between galaxies allows
us to propose that its formation is the result of compression
of colliding gas flows of galaxies in contact.

\paragraph{{\bf 2. System Arp~194}}

Arp~194 is a system of recently collided galaxies, where the  southern galaxy (S) passed through the gaseous disc of the  northern galaxy (N) which in turn consists of two close components (Fig.2). This system is of  special interest due to the presence of regions of active star-formation in the bridge between galaxies, the brightest of which (the region A) has a size of at least 4 kpc.  We obtained three spectral slices of the system for different slit positions 
and estimated the radial distribution of line-of-sight velocity and velocity dispersion as well as the intensities of emission lines and oxygen abundance $12+\log(\mathrm{O/H})$.  A gas in the bridge is only partially mixed chemically and spatially: we observe the O/H gradient along the galactocentric distances both from the centers of S and N galaxies, and a high dispersion of O/H in the outskirts of N-galaxy. Velocity dispersion of the emission-line gas has the lowest values in  the star-forming sites in the  bridge and exceeds 50-70 km/s in the disturbed region where there are no bright extended star-forming regions, and, judging from the line ratios, the emission belongs to a diffuse ionized gas, with collisional mechanism of excitation of atoms.

Based on the comparison of stellar evolutionary models with the SDSS photometrical data, and using our kinematical profiles, we measured the ages and masses of stars and the dynamical masses of individual foci of star formation.  We confirm that the largest emission island (region A) is most probably a gravitationally bound TDG with the age of $10^7 - 10^8 $ yr.  There is no evidence of the significant amount of dark matter in this dwarf galaxy. The measured velocities of gas along the slits crossing TDG agree with the scenario where this young star-forming island and its surrounding gas falls into the disc of S-galaxy. 
\begin{figure}
\vspace{-6.5cm}
\centering
			\includegraphics[width=0.9\columnwidth]{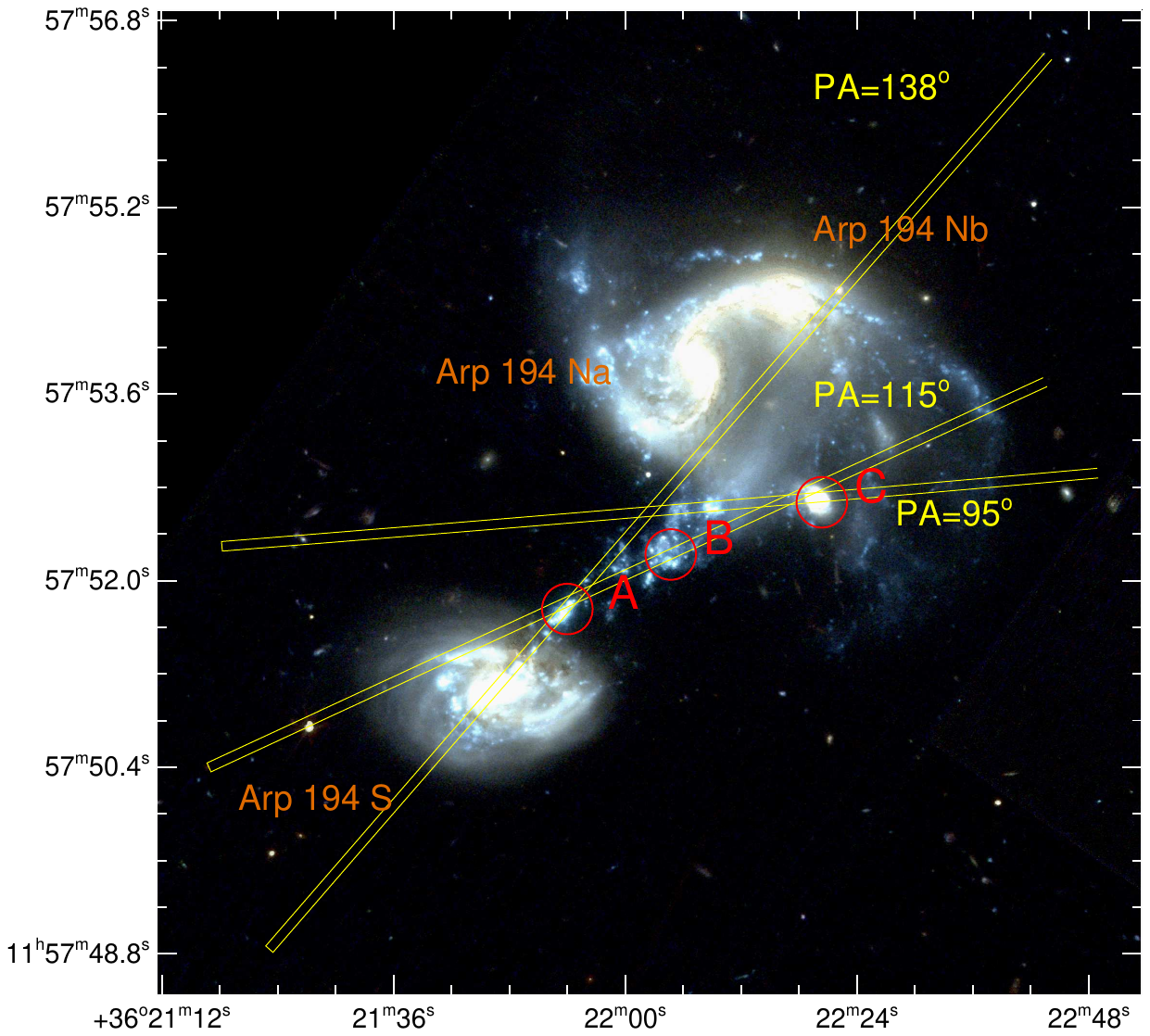}
			\vspace{-4.0cm}
		\caption{The HST composite colour image of Arp 194 in BVI bands with overplotted positions of the slits.  }
	\end{figure}
\paragraph{{\bf 3. System Arp~305}}

Arp~305 is a small group of galaxies dominated by the wide pair of  interacting spiral galaxies of moderate luminosity: NGC~4016 and NGC~4017.  For the adopted distance 50 Mpc the projected separation of galaxies is 86 kpc. The largest concentration of HI between galaxies coincides with the optical chain of blue knots of  total magnitude $B\approx 17$ ,  immersed in a faint haze, most clearly visible in UV  \citep[see][]{Sengupta2017}. It is located about halfway between the galaxies and stretched  along the line connecting them at least at  7 kpc. This object is considered as the TDG candidate. Simple Stellar Population models applied to the photometric data demonstrated that 
the  separate clumps have a very small age, which allows to conclude that this observed stellar island was recently formed from the local concentration of gas in the tidal tail. 

\begin{figure}
\centering
	\includegraphics[width=0.4\columnwidth]{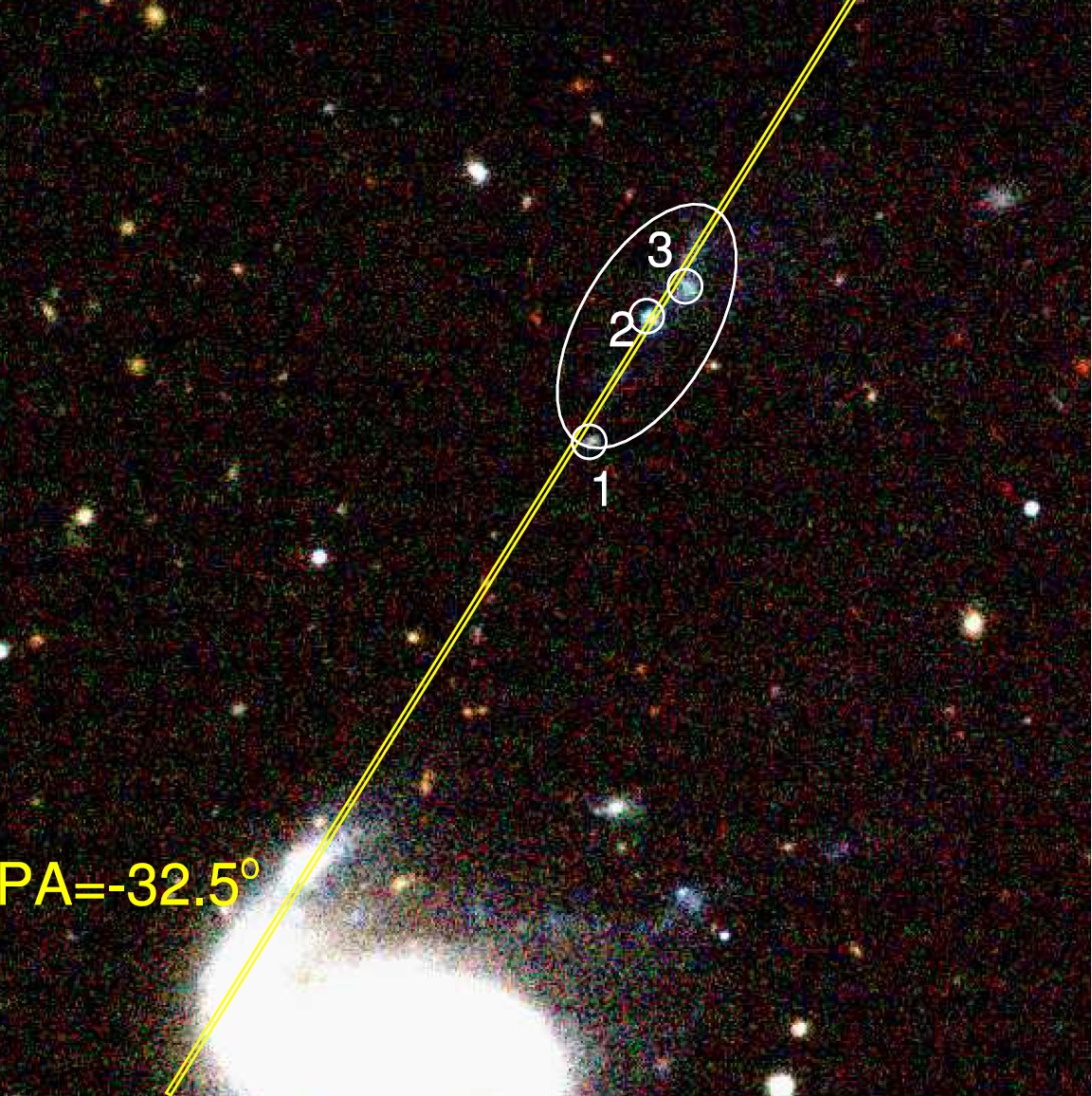}
	\includegraphics[width=0.298\columnwidth]{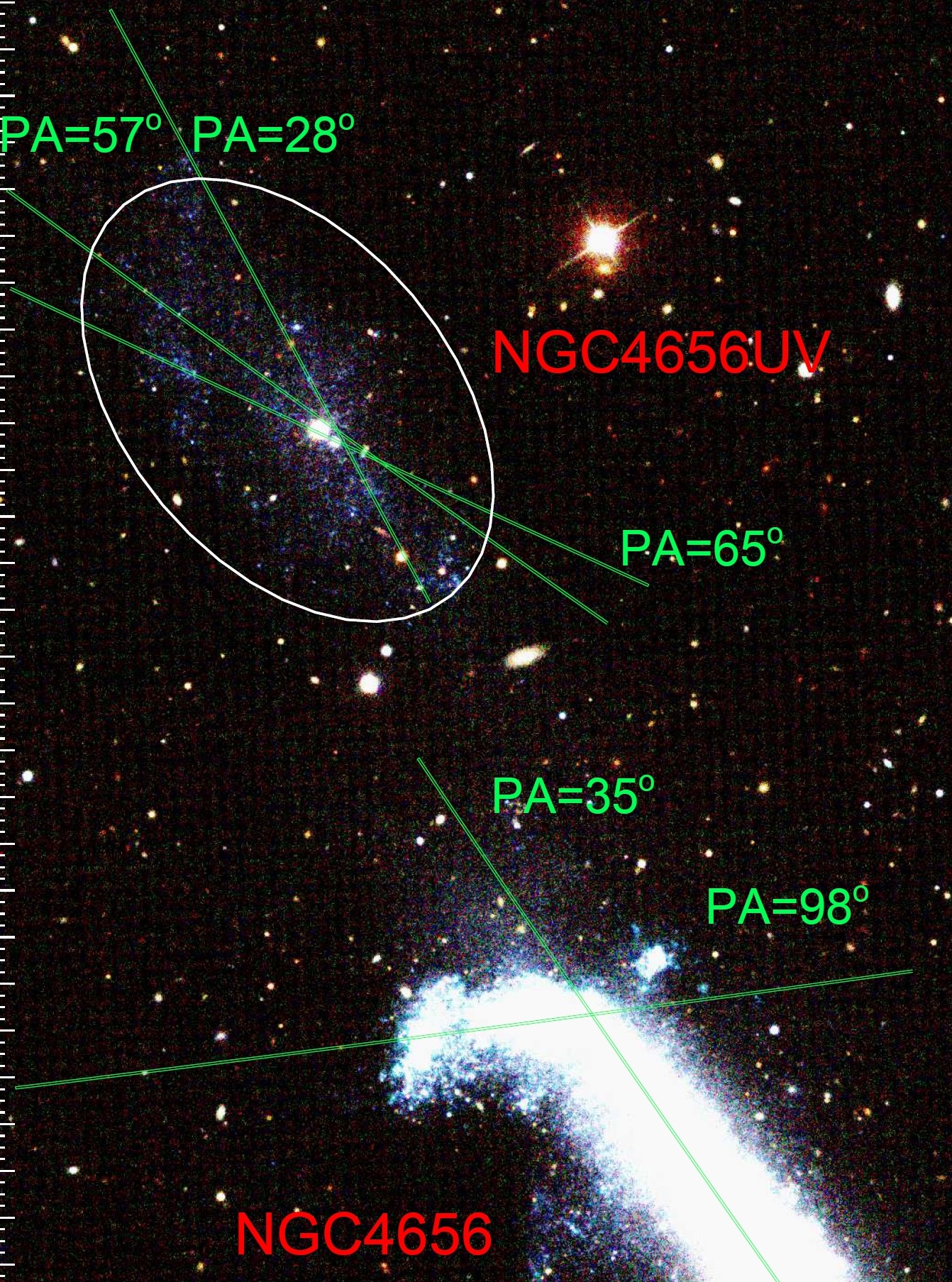}
		\caption{The positions of the slits overplotted on the {\it g,r,i}-band image of Arp~305 (right) and NGC~4656 (left). \label{fig3}  }
	\end{figure}

This object was observed with the slit position shown  at composite {\it g,r,i}-image in Fig. \ref{fig3}.  We found that gas velocities in  TDG  are more disturbed than in the adjacent regions.  At the same time, we observed neither noticeable rotation nor expansion of TDG:  mean velocities of its two brightest clumps differ by no more than 20 km/s. The oxygen abundance of TDG
found by izi \citep{Blanc2015} and O3N2 \citep{Marino2013} methods is 8.1 -- 8.3, which is too high for its absolute magnitude $M_B\sim -16$ mag.

The rough estimate  of dynamical mass ($M_{dyn}= (2\pm 1)\cdot 10^9 M_\odot$)  of TDG is much higher than its stellar mass, however it is 
comparable with the observed mass of HI connected with it.    The fate of the TDG in Arp~305 seems similar to that of the stellar
island in Arp~194 discussed above – it will be soon accreted by the parent galaxy.

A column gas density in the tidal bridge is maximal in the region of TDG, reaching  $4\times10^{20}$ g cm$^{-2}$ \citep[see][]{Sengupta2017}.  A rough estimate of Jeans mass for the observed velocity dispersion of gas  is about $6\cdot10^8 M_\odot$, which agrees with the total (mostly gaseous) mass of TDG, as it is expected if the mass of TDG is close to that required for gravitational bounding.  Note however that the free-fall time for this density is too long -- about $10^8$ yr, being comparable to the time of dynamic evolution of the perturbed galaxies. Hence, either a gas which fills TDG was inhomogeneous initially, or the current star formation was triggered by some external pressure.

\paragraph{{\bf 4. UV satellite of NGC~4656}}

The unusual  dwarf low surface brightness  galaxy NGC~4656UV is a close satellite of the edge-on galaxy NGC~4656. It  is barely visible in the SDSS image, however it  has the the enhanced brightness in the UV range.  
Multiwavelength  data analysis  carried out by \cite{Schechtman-Rook2012}  led these authors to the conclusions  that  NGC~4646UV is a low-metallicity TDG candidate whose major burst of star formation occurred within the last $\sim$ 260-290 Myr. 

We did several spectral cuts across both galaxies (see Fig.\ref{fig3},  right panel) to obtain  the  kinematical parameters and gas-phase metallicity of NGC~4656UV and NGC~4656. Our estimates of emission gas velocities of NGC~4656UV parallel with photometrical and HI data speak in favour of this system to be a dark matter-dominated gravitationally bound axisymmetric galaxy with the dynamic mass of about $2\cdot10^9$ solar masses, rather than a  tidal dwarf candidate. It has  quite axisymmetric exponential disc with the low central surface brightness $(\mu_{0})_r$= 24.1 ~mag/arcsec$^2$  and the radial scale length of about 2 kpc. The parameters of NGC~4656UV are close to that observed for separate  ultra-diffuse galaxies.
 
  Oxygen gas-phase abundances found for the brightest HII-region of the UV dwarf, as well as for the emission gas of the main galaxy NGC~4656 at the side facing UV dwarf,  are equally low: 12+log O/H = 7.8 - 8.1 (\textit{izi}-method). Both the low abundance and non-circular gas motions in NGC~4656 parallel with the observed young stellar population of NGC~4656UV give evidences of the current accretion of metal-poor gas on the discs of these galaxies due to tidal interaction between NGC4656 and NGC4631. Gas flows between these two galaxies are clearly seen at the HI map of the interacting system \citep{Rand1994}. A current star formation  induced by the accreted gas may explain the enhanced UV brightness of the  `ghost'  LSB-galaxy, which would be nearly invisible otherwise.
\paragraph{{\bf 5. NGC~90, a member of Arp~65 system}}

A peculiar galaxy NGC~90 is a pair member of interacting system Arp~65 (NGC~90/93), where both galaxies have a comparable luminositiy. NGC~90 possesses a pair of unusually thin and regular inner spiral arms of Grand Design type which straighten at the periphery, passing into the  short tidal tails. Curiously, the line-of-sight  (LOS)  velocities of the  paired galaxies differ by more than  400 km/s.  Such velocity discrepancy is unusually high for  interacting galaxies with tidal structures. Both galaxies are the members of a group SRGb063  where X-ray gas was detected \citep{Mahdavi2004}.

\citet{Sengupta2015} found  that about a half of hydrogen mass of NGC~90  looks strongly displaced with respect to its  main body – both in space  and in the velocity field. High-speed gas looks like a  giant non-rotating massive HI cloud, comparable with the optical galaxy by its size, which is projected on the  SE part of the disc of NGC~90.  The LOS velocity of this HI cloud does not agree with the velocity of the  galaxy, exceeding its central velocity at about 340 km/s, so the cloud may hardly be gravitationally linked with the galaxy. It makes this system  very interesting for further investigation.
 \begin{figure}
 \centering

			\includegraphics[width=0.6\columnwidth]{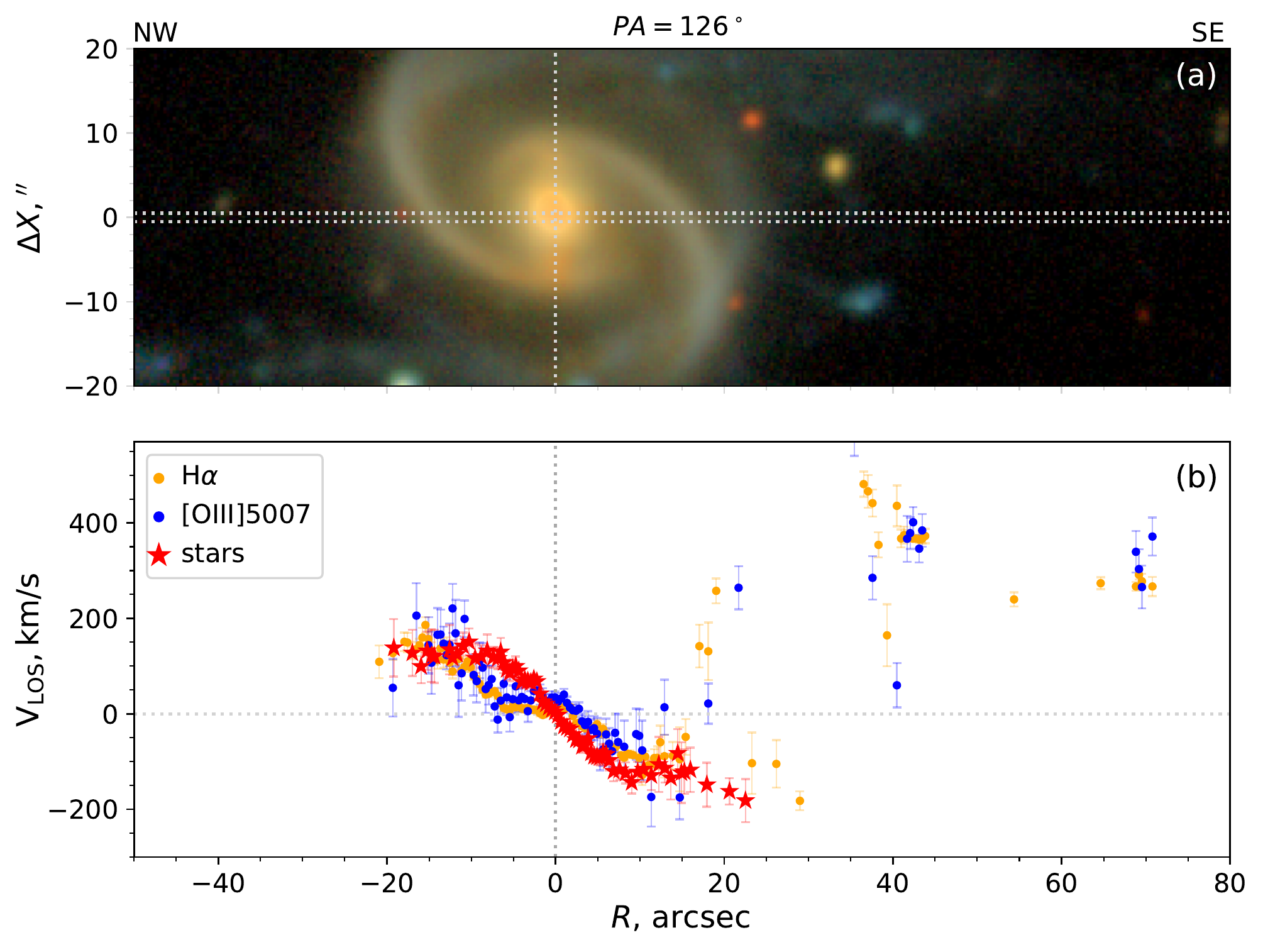}
		\caption{Profiles of stellar and ionized gas velocities for NGC~90 along the slit passing through its center and the HI cloud observed to the right of the galaxy.  }
	\end{figure}
 
 We fulfilled two parallel spectral cuts with  the orientations of both slits in the NW-SE direction. One of them crossed  the centre of the galaxy, and another one was directed along the northern spiral arm.  The first cut also run through the  position of high velocity HI cloud. The most interesting result is that we found the emission patches beyond spiral galaxy, in the area of this peculiar cloud  (see Fig.4). Their velocities change along the slit, increasing up to the values corresponding to that for HI  cloud. It means that there exists a chain of HII regions between the galaxy and the cloud, and
 there are local regions of star formation  which evidently are detached from the galaxy disc.

 We argue that what we observe in this galaxy  is similar  to the HI + $H_\alpha$ tails of so-called jelly-fish galaxies \citep[as an example, see][]{Ramatsoku2019}. In this case it is a ram pressure of the intra-group hot gas which is responsible for sweeping gas out of the disc.  
 
Indeed, the difference between the observed velocity of NGC~90 and the mean velocity of the group which it belongs to, is as high as  570 km/s. The gaseous tail  formed as the result of sweeping of gas, is apparently elongated along the line of sight, which creates the illusion of a cloud with a high column density of gas projecting onto the galaxy region. Local emission regions we found in this tail evidences that star formation beyond galaxy disc, inspired by the ram pressure of gas, has not completely faded yet.  

\paragraph{{\bf Acknowledgment}}
This work was supported by RFBR grant 18-02-00310 and 18-32-20120.

\nopagebreak
\begingroup
\let\clearpage\relax
\bibliographystyle{aa}
\nopagebreak
\bibliography{template}
\endgroup
\end{document}